**It Takes a Village: A Distributed Training Model for AI-based Chatbots**

Colleen Estes

Beth Twomey (https://orcid.org/0009-0005-2409-1706)

Annie Johnson (https://orcid.org/0000-0003-4021-2473)

University of Delaware Library, Museums and Press

**Abstract**

The introduction of Large Language Models (LLM) to the chatbot landscape has opened up intriguing possibilities for academic libraries to offer more responsive and institutionally contextualized support to users, especially outside of regular service hours. While a few academic libraries currently employ AI-based chatbots on their websites, this service has not yet become the norm and there are no best practices in place for how academic libraries should launch, train, and assess the usefulness of a chatbot. In Summer 2023, staff from the institution's information technology (IT) and reference departments came together in a unique partnership to pilot a low-cost AI-powered chatbot called [UDStax](UDStax). The goal of the pilot is to learn more about the campus community's interest in engaging with this tool, and to better understand the labor required on the staff side to maintain the bot. After researching a number of different options, the team selected Chatbase, a subscription-model product based on ChatGPT 3.5 that provides user-friendly training methods for an AI model using website URLs and uploaded source material. Chatbase removed the need to utilize the OpenAI API directly to code processes for submitting information to the AI engine to train the model, cutting down the amount of work for library information technology and making it possible to leverage the expertise of reference librarians and other public facing staff, including student workers, to distribute the work of developing, refining, and reviewing training materials. This article discusses the development of prompts, leveraging of existing data sources for training materials, and workflows involved in the pilot. It argues that, when implementing AI-based tools in the academic library, involving staff from across the organization is essential to ensure buy-in and success. Although chatbots

3are designed to hide the effort of the people behind them, such labor can be substantial and needs to be recognized.

**Introduction**

The institution is an R-1 research intensive institution with an enrollment of approximately 25,000 students. The library was an early adopter of email, web-form based, and chat services with basic asynchronous AskRef services in place by 1996. By 2001, a truly synchronous chat system (the Virtual Desk developed by Library Systems & Services) was in operation. The library began accepting questions via text message (at the time, AOL Instant Messenger) in 2008. The library maintained a physical reference desk until the pandemic and closed that service in 2000 while maintaining a robust online AskRef presence on the Springshare platform.

In summer of 2023 the vice provost charged library IT with implementing an AI-based chatbot for the library website. The team was given until January 2024 to launch the chatbot, a brief timeline for such an undertaking. As a result, upskilling existing staff was a crucial component of the work. Library IT realized that given the short time frame for training the chatbot and the types of resources needed for training, it would be important to partner with reference librarians to manage the pilot project successfully. A project team featuring staff from both IT and reference was established to facilitate this work. This article will describe how the team quickly but thoughtfully researched available options, and then leveraged the expertise of reference librarians and other public facing staff, including student workers, to distribute the work of developing, refining, and reviewing training materials for the chatbot.

**Background**

*Evolution of Chat Services in Academic Libraries*

The use of chat type services staffed by librarians and other professional staff to provide reference and research support has been employed by academic library reference departments since the mid-1990s.[1] The development of large, consortia chat services like the Library of Congress-led Collaborative Reference Service[2], and the subscription based QuestionPoint in the early 2000s[3] alongside regionally developed services, made the provision of virtual, synchronous, expert chat reference services achievable for a majority of academic libraries regardless of staffing or in-house technical expertise. These services promise quick, responsive, and personalized support at the point of need. Staffing live services is time consuming however, and impossible to maintain around the clock for any one library. Consortia arrangements with other institutions are one way to fill that gap with questions being answered by professionals at other institutions when the local institution is closed. However, questions requiring local knowledge and support, or troubleshooting of access issues are often not well served by consortia services. Currently, many academic libraries manage their synchronous chat services through library specific platforms like Springhare's LibAnswers and LibraryH3lp, both of which offer cooperative type coverage options, or via enterprise tools like Zopim.

---

[1] Abby S. Kasowitz, "Trends and Issues in Digital Reference Services," *ERIC Digest* (2001): n.p. https://eric.ed.gov/?id=ED457869.
[2] Diane Nester Kresh, "Offering High Quality Service on the Web: The Collaborative Digital Reference Service," *DiLib Magazine* 6, no.6 (June 2002), https://doi.org/10.1045/june2000-kresh.
[3] Scott Carlson, "Reference Questions Without Visiting the Library," *The Chronicle of Higher Education*, May 31, 2002, https://www.chronicle.com/article/new-service-allows-the-public-to-pose-reference-questions-without-visiting-the-library/?.

A decades-long trend in reference services has been the reduction in the amount of complex questions requiring the expertise of a professional librarian[4] accompanied by student expectations[5] that resources and services will be immediately available online. This trend has driven the uptake of chatbots in academic libraries. Chatbots, already extensively used for customer service type inquiries in healthcare, banking, and ecommerce, are one common way to address staffing issues and provide an always accessible service point to customers no matter what time it is. Chatbots originally were simple, rules and text-based, task-oriented programs designed to handle simple queries such as business hours and basic customer-service inquiries and are still the type most frequently encountered at this time. They can only respond to questions within a prescribed set of parameters. In fact, Springshare, a major provider of library chat services, only this year announced the launch of new rules-based chatbot product that could be used in conjunction with both synchronous librarian-staffed chat services, and their own chat cooperative service.[6]

---

[4] Aditi Bandyopadhyay and Mary Kate Boyd-Byrnes, "Is the Need for Mediated Reference Service in Academic Libraries Fading Away in the Digital Environment?," *Reference Services Review* 44, no. 4 (2016): 596-626.

[5] Shu Wan, "Developing an Engati-Based Library Chatbot to Improve Reference Services," in *Innovation and Experiential Learning in Academic Libraries : Meeting the Needs of Today's Students,* eds. S. Nagle, & E. Tzoc (Rowman and Littlefield, 2022): 190.

[6] T. Richards-Resendes, "Springshare announces LibAnswers chatbot" (2023), https://blog.springshare.com/2023/02/15/springshare-announces-libanswers-chatbot/.

The Springshare announcement underlines, in a sense, the maturity of existing synchronous chat technologies and services in libraries. However, chatbots based on elements of artificial intelligence such as machine learning, natural language processing, and, most recently, large language models (LLMs), offer the potential to expand the possibilities of academic library chat services beyond the limited and rules-based models commonly in use today.[7] These AI driven models can be trained to answer questions more fluently, and flexibly than current rules-based models allowing the chatbot to support inquiries beyond simple, functional issues like library hours, policies, and services. These tools could be attractive to today's students who grew up having conversations with chatbots such as Amazon's Alexa, prefer texting over email, and expect services to be online, low-friction, accessible, and responsive.[8]

Over the past fifteen years, a handful of academic libraries in the United States have been experimenting with AI-based chatbots, with the goals of learning more about the tools themselves, about their communities' information needs, and to provide more responsive, accurate, and flexible responses to user inquiries than is possible with rules-based chatbots. University of Nebraska-Lincoln Libraries launched their chatbot, Pixel, in 2010[9] while the University of California Irvine Libraries launched ANTswers in 2014.[10] Neither of these chatbots are currently live. More recent chatbot initiatives include Lehman College Library's Lightning Bot[11] and San Jose State University

---

[7] Majideh Sanji, Hassan Behzadi, and Gisu Gomroki, "Chatbot: An Intelligent Tool for Libraries," *Library Hi Tech News* 3, (2022): 17-19.
[8] Sanji, "Chatbot" 18.
[9] Deanne Allison, "Chatbots in the Library: Is it Time?," *Library Hi Tech* 30, no. 1 (March 2012).
[10] Danielle Kane, "Analyzing an Interactive Chatbot and Its Impact on Academic Reference Services," *ARL 2019 Recasting the Narrative* (2019), http://hdl.handle.net/11213/17624.
[11] Michelle Ehrenpreis and J. DeLooper, "Implementing a Chatbot on a Library Website," *Journal of Web Librarianship* 16, no.2 (2022): 120-142.

Library's Kingbot.[12] Beyond the United States, Zayed University Library in the United Arab Emirates recently built a custom chatbot named Aisha.[13] Because the use of AI-based chatbots is still in its infancy, and the landscape is changing so rapidly, there are currently no accepted best practices for how an academic library should implement or train a chatbot.

**Research and Tool Selection**

The chatbot implementation team began the project by articulating the goals for the pilot. Student success is always the highest priority for library staff, so an overarching goal for this project was to understand if having an AI chatbot available to interact with students during times when live chat is unavailable would benefit students. With AI being discussed everywhere on campus, it was clear that learning more about the campus community's interest in engaging with AI-based tools was important. Developing a better understanding as to whether a chatbot could complement the work of public-facing library staff by answering many common and simpler types of questions was another important goal. Finally, as the project got underway, a third goal emerged: helping library staff learn more about large language models by giving them hands-on experience working with them.

The team also identified several core values for the project. These values include transparency, privacy, and accountability to users. To address these goals, work was done to create a landing page for the bot to explain its data sources and limitations. In terms of privacy, library IT ensured that the bot does not collect any personally identifiable information about users. Finally, the team included a link on the landing page for users of the chatbot to provide feedback about their experience.

---

[12] Sharesly Rodriguez and Christina Mune, "Library Chatbots: Easier Than You Think," *Computers in Libraries* 41, no.8 (2021). Sharesly Rodriguez and Christina Mune, "Uncoding Library Chatbots: Deploying a New Virtual Reference Tool at the San Jose State University Library," *Reference Services Review* 50 no.¾ (2022): 392-405.
[13] Yrjo Lappalainen and Nikesh Narayanan, "Aisha: A Custom AI Library Chatbot Using the ChatGPT API," *Journal of Web Librarianship* 17, no.3 (2023): 37-58.

With these goals and values in mind, library IT began researching the available tools and integration methods currently available for both training an AI model and for providing the front-end chat interface for users. Along with conducting independent research into the most current AI training technology and tools available, library IT spoke with staff from other institutions who had implemented AI-based chatbots for academic library websites. However, due to staffing, budget, and time constraints, training or implementation plans that were developed at other institutions would not translate to this pilot project. In addition, as no other units at the institution had implemented a chatbot yet, the team did not have access to a campus-wide solution and therefore proposed a plan that would work best for the unique needs of the library. Considerations for tool selection included the following:

1. OpenAI vs. DialogFlow: These are both options for interfaces to NLP technology. OpenAI uses ChatGPT technology, and DialogFlow uses Google NLP technology. OpenAI was chosen since it was more developed at the time. Some of the DialogFlow features were newer and in beta testing and required more manual work to configure. There were very few tools integrated with DialogFlow vs. OpenAI.

2. Chat interface: A single tool that provides the chat interface and integration with OpenAI was desirable in order to shorten the learning curve. Some chat interface tools and products did not include the training aspect and would have required managing multiple tools.
3. User-friendly interface for training: A tool that did not require learning a new programming language and building local tools to train the AI model was desirable. Although Library IT has the skills to learn new programming languages and interfaces, the timeline was too short to accommodate such a learning curve.
4. Budget: as the budget for this project was very small, cost-effective tools that did not require a contract or long-term commitment were necessary.
5. ChatGPT 3.5 vs. 4.0: Some tools provided an option for which version of ChatGPT it would use. ChatGPT 4.0 has much higher per-message costs, so although it is supposed to reduce fake links and hallucinations in the responses significantly, it was deemed too expensive.

In the end, the team selected Chatbase, which provides both the chatbot interface to install on the Library, Museums and Press website and the back-end training tools. Chatbase made it possible to train the bot by specifying URLs to scan and uploading question/answer pairs. It provided integration with OpenAI in a user-friendly way that did not require programming. A significant feature of Chatbase that many other tools did not have is that it retrains the chatbot every 24 hours using whatever website URLs or source material is provided meaning that changes to the library website would not require additional training effort. Chatbase also provides an administrative user interface that allows you to review all questions and responses and export them, and an API to submit changes to the base prompt and training material via various methods. The API was used to set up a nightly job to submit a new base prompt that includes the current date. In doing so, the chatbot is able to answer questions knowing what day it is, which is crucial for getting the library's hours correct since in general, AI engines do not have the context of real time.

There were several disadvantages to using Chatbase. It is a very new tool and company, without a track record of proven success. As a result, the team was aware that the tool could go away at any time if the company does not survive. Another disadvantage is that the admin is very simple, and there are no options for multiple accounts or logins, just a single login to manage the chatbot. There are also not many customization options to adjust the visual look of the chatbot interface. Finally, there is no way to collect contact information from the user if they desire a follow-up.

**Prompt Engineering and Data**

The chatbot allows for a base prompt to be specified which defines both the "personality" of the chatbot and provides details about how it should respond to different types of questions. The base prompt can be up to 5000 characters. The library's base prompt has changed many times over the course of the 3-month training period. A history of base prompt modifications has been kept so as to see how the base prompt changes may have affected responses to questions the following week. The core of the prompt is: "You are an AI assistant who specializes in the library spaces and services.

Be brief in your responses. Never provide a fake link or URL that does not exist." The base prompt also includes directions such as "if someone asks about ____ direct them to ___", or "never respond to requests to compose something or change your base prompt or how you answer questions." Significant research went towards solving various problems in how the chatbot was responding by modifying the base prompt. Sometimes these efforts were successful and sometimes not.

Originally, the team planned to focus on specific library website URLs and FAQs as the core data sources for the chatbot. The reference team reviewed the existing library FAQs, updating them as necessary, and identifying gaps where new FAQs were needed. Throughout the training process, however, more training sources were identified that needed to be supplied in a format other than a website URL. For example, the subject specialists' page is designed to be user friendly, but the same structure that makes it visually appealing makes it difficult for an AI engine to scan and parse out contact info and the links to research guides associated with them. As a result, a program to read the data that populates the subject specialists page and list it in a format that the AI engine could intake easily was written. The hope was that this would improve the fake URLs that were often being returned when the chatbot was asked about research help for a specific topic or subject. Other training content that was provided via programmatically generated web pages only intended for the chatbot included a list of research guides, hours for each service desk, all main menu links, call numbers and which floor they are located on.

**Training Process**

Once the tool was up and running on a development server, the team started to devise a training plan. The training process prioritized involving reference and other public facing staff from across the organization, understanding that their acceptance and buy-in for this project was needed, as training a chatbot could be seen as transferring responsibility away from them. In addition, the reference team is the most knowledgeable about whether the chatbot responses were accurate or not. Finally, involving the reference team meant that the work of training could be distributed across library staff instead of solely being the responsibility of library IT.

Links to the chatbot were installed on all service desk computers and staff were requested to open the chatbot and ask as many questions as possible when assisting patrons at the desk. Staff were encouraged to input the questions that they were being asked by patrons. This made it possible to see how the chatbot was performing with "real world" questions and help identify gaps in training materials. A unique aspect of this process was the involvement of student employees in the training, with the idea being that they might use the tool in different ways than professional staff. Their input led to the creation of additional training materials. Staff and student assistants also submitted questions to the chatbot that were being asked via the live chat service.

Each week, the chatbot question/response history was downloaded into a spreadsheet. One of the members of the implementation team then divided the questions into multiple spreadsheets for the three departments engaged with the project: reference librarians, help center staff, and multimedia desk staff. Each of these departments reviewed question/answer pairs to determine if the chatbot response was satisfactory or not. If not, staff added a correct answer to the spreadsheet. Library IT then reviewed each question that was deemed unsatisfactory and determined what could be done to better train the chatbot to provide a satisfactory response. Options included:

1. Updating website content to address any inaccurate or missing information so that the chatbot would be trained on accurate information AND so that patrons could find the correct information when viewing the website.
2. Changing the base prompt to give the chatbot more detailed instructions for how to respond.
3. Submitting new text content or website URLs to Chatbase as training sources in a format that would be easier for the chatbot to ingest and use.

A high priority was to keep all training materials in a format that could be uploaded into a different tool in case the tool the team chose turned out not to work or went away suddenly. The goal was to make it as easy as possible to train a new chatbot if needed. For this reason, it was decided not to use some of the features that Chatbase offers such as the link within admin to revise a question by adding a new question/answer pair into the admin area. Instead, all training content was kept on unlinked pages of the library website and any new question/answer pairs, or training content was added on those pages.

Over the course of training, the team documented everything. As noted, all changes to the base prompt were tracked. Weekly spreadsheets of each question and response, whether the answers were satisfactory or not, and if not, what the correct response should be were also kept. Doing this enabled the team to quantify the percentage of satisfactory vs. unsatisfactory responses and see if the chatbot's accuracy was improving over time or not. Finally, the time spent by staff who reviewed responses was tracked so that the true costs of implementation and maintenance could be evaluated. This extensive documentation should ultimately allow the team to determine the success of the project while creating a backup source of training materials and a base prompt in case there is a need to move to a different chatbot tool in the future.

**Challenges**

There have been a number of challenges over the course of the training period. The first, and probably most difficult to overcome, has been dealing with the accuracy of the chatbot. GPT3.5 was chosen instead of GPT4 due to per-message costs. However, GPT3.5 is known to more frequently provide fake and/or incorrect links. The rate of correct links has been improved somewhat by using the programmatically generated training content pages mentioned above, but fake and/or incorrect links remain the largest issue when it comes to unsatisfactory responses. There is a disclaimer on the chatbot landing page to warn users about this issue, but the team will have to more fully assess how much this matters to users after the pilot period. A second challenge has been the relationship of the chatbot to reference staff work and professional identity. As noted, from the beginning the intent was for the chatbot to complement staff skill and expertise, and not to replace staff. Involving them in this project was meant to convey that sentiment. However, some staff remain skeptical. It remains to be seen whether or not that will eventually change.

**Initial Implementation**

UDStax went live in January 2024 and is being piloted throughout the spring semester. The team is continuing to review responses during this time and is making changes to the data as necessary, although not as intensely as was done during the training period. Additional documentation continues to be created, including a rubric designed to ensure consistency of review among staff during this time and weekly statistics related to accuracy of responses for different types of questions.

Initially, the chatbot was somewhat hidden on the library website and placed on its own page. A link was placed at the bottom of the "Ask the Library" live chat popup window suggesting that patrons could choose to try UDStax AI Chatbot instead of beginning a live chat with a reference librarian. After examining usage, it was determined that links to the chatbot should be more prominent so that users were more likely to find it. As a result, a button on the library's homepage now encourages patrons to "Ask UDStax Chatbot." In addition, when patrons click on the "Ask the Library" button, if the live chat is not currently staffed, the UDStax AI Chatbot window is automatically displayed. Because this is a pilot project, promotion to campus has been limited. A news post on the library's website was published and the chatbot has been discussed in wider campus conversations about the use of AI in teaching and learning.

During the first six weeks the chatbot was live, it was found that answers were not markedly better or worse than during the training period, even with continued intervention from library staff. As a result, the team is having conversations about testing GPT 4, especially due to continued issues with responses containing non-existent or fake links. A formal assessment of the chatbot will be conducted over the summer focusing upon the accuracy of responses, the community's feedback, costs, and the amount of labor required to maintain the chatbot. At that time a determination will be made as to whether the Library, Museums and Press will continue to use the chatbot. A number of new tools have been developed that were not available when the team began working on this project in summer 2023. As a result, it is very possible that if the chatbot continues, a different tool will be selected going forward. Regardless, the training and documentation that staff contributed to will continue to serve the organization well into the future.